\RequirePackage{fix-cm}
\documentclass[smallcondensed]{svjour3}     
\smartqed  
\usepackage{graphicx}
\journalname{Hyperfine Interactions}
\begin{document}

\title{Stringent tests of QED using highly charged ions
}


\author{V.~M.~Shabaev$^{1}$         \and
  A.~I.~Bondarev$^{1,2}$     \and
  D.~A.~Glazov$^{1}$    \and
  M.~Y.~Kaygorodov$^{1}$    \and
  Y.~S.~Kozhedub$^{1}$    \and
   I.~A.~Maltsev$^{1}$     \and
  A.~V.~Malyshev$^{1}$     \and
  R.~V.~Popov$^{1}$              \and
  I.~I.~Tupitsyn$^{1}$    \and
  N.~A.~Zubova$^{1}$  
}


\institute{V.M. Shabaev \\
   \email{v.shabaev@spbu.ru}\\
 $^1$ Department of Physics, St. Petersburg State University, Universitetskaya 7/9,
199034 St. Petersburg, Russia  \\
$^2$ Center for Advanced Studies, Peter the Great St. Petersburg Polytechnic University, 
195251 St. Petersburg, Russia  
}
\date{Received: date / Accepted: date}

\maketitle

\begin{abstract}
  The present status of tests of QED with highly charged
  ions is reviewed. The theoretical predictions for the Lamb shift and the transition energies
  are compared with
  available experimental data. Recent achievements in studies of the hyperfine splitting
  and the  $g$-factor isotope shift with highly charged ions are reported.
  Special attention is paid to  tests of QED within and beyond
  the Furry picture at strong-coupling regime.  Prospects for tests of QED at
  supercritical fields that can be created in low-energy heavy-ion collisions
  are discussed as well.
\end{abstract}

\section{Introduction}
\label{intro}
For almost four decades since the creation of quantum electrodynamics (QED)
in its present form, 
 tests of QED were mainly restricted by light atomic systems:
hydrogen, helium, positronium, muonium, etc. The calculations of these systems are generally based
on expansions in two small parameters: $\alpha$ and $\alpha Z$, where  $\alpha$ is the
fine structure constant and $Z$ is the nuclear charge number. So, the studies of these
systems provided tests of QED in few lowest orders in $\alpha $  and $\alpha Z$ only.
A possibility to extend this rather small region of the QED tests appeared in the late 1980s
 when first high-precision measurements with heavy few-electron ions were performed.
In contrast to light atoms, the parameter  $\alpha Z$ in highly charged ions is no longer small. 
Therefore, the calculations of these ions must be performed to all orders in  $\alpha Z$.
From one side, it provides us with serious technical and, in some cases, even conceptual problems
but, from the other side, it gives us a unique opportunity to test QED in a new region:
nonperturbative in  $\alpha Z$ regime.

The basic idea of the theoretical approach to the QED calculations of highly charged
few-electron ions can be formulated as follows. Since the number of electrons in
these ions is much smaller than the nuclear charge number, to zeroth order we can take into
account only the interaction of the electrons with the Coulomb field of the nucleus, $V_{\rm C}(r)$,
and neglect the interaction of the electrons with each other. This means that to
zeroth order the electrons obey the Dirac equation ($\hbar=c=1$),
\begin{equation}
( -i \, \vec{\alpha} \cdot \vec{\nabla} +m\beta+ V_{\rm C}(r) 
\, )\, \psi \, (\vec{r})
= \, E \, \psi \, (\vec{r})\,. 
\end{equation}
Then, the interelectronic-interaction and QED effects are accounted for by perturbation theory
in the parameters $1/Z$ and $\alpha$, respectively. Due to smallness of these parameters,
precise QED calculations of these systems are definitely possible. Moreover, for very
heavy ions the parameter $1/Z$ becomes comparable with $\alpha$ and, therefore, 
 all contributions can be classified by the parameter $\alpha$. This perturbation theory
can be further improved by incorporating into the Dirac equation a screening potential, $V_{\rm scr}(r)$,
which partly accounts for the electron-electron interaction effects. Then, in the higher
orders, to avoid double counting of some electron-electron interaction effects, one should add the interaction
with the potential $-V_{\rm scr}(r)$. These two schemes of the QED perturbation approach
are known as the Furry and the extended Furry picture, respectively.

\section{Lamb shift and transition energies}
\label{sec:1}
In Table~\ref{tab:1} we present the theoretical contributions to the $1s$ state
Lamb shift in H-like uranium \cite{gla11}. The Lamb shift is defined as the difference between
the exact total energy and the energy which is derived from the Dirac equation for the
point-charge nucleus. The finite nuclear size effect was evaluated for the Fermi
model of the nuclear charge distribution,
including the nuclear deformation effect \cite{koz08}.
The Furry picture of QED incorporates the contributions
of the first order in  $\alpha$ (one-loop QED) \cite{yer15}
and the second-order in  $\alpha$
(two-loop QED) \cite{yer03}.  The calculation of the nuclear recoil effect \cite{sha98} 
requires using QED beyond the Furry picture at strong-coupling regime.
Finally, one should account for the nuclear-polarization effect, which in terms of the
Feynman diagrams is described by two-photon exchange between the electron and the nucleus
with excited intermediate nuclear states \cite{plu95,nef96}. As one can see from Table~\ref{tab:1},
the total theoretical Lamb shift value agrees well with the experimental one \cite{gum05} but
has ten times better accuracy. This provides a test of QED at strong field on a 2\% level. 

A higher accuracy was achieved in experiments with heavy Li-like ions \cite{sch91,bra03,bei05}.
The comparison  of the most precise experiment on the $2p_{1/2}-2s$  transition energy
in Li-like uranium \cite{bei05} with the related theory 
\cite{gla11,koz08,yer07,sap11}  provides a test of strong-field QED on a 0.2\% level.

In Ref. \cite{cha12}, on the grounds of new  measurements with heliumlike titanium ($Z= 22$) and a statistical
analysis of all the data available in literature,  it was claimed that there exists
a systematic discrepancy between theory and experiment, which scales approximately as $Z^3$.
Although the experimental results of a number of works
(see Refs. \cite{tra09,ama12,kub14,bei15,epp15,mac18})
for the transition energies in He-like ions agree with the most elaborated QED calculations \cite{art05},
independent theoretical predictions would be very desirable. In Table~\ref{tab:2} we present the results
of our calculation of the $1s2p\,^1P_1-1s^2\,^1S_0$  transition energy in heliumlike titanium and
compare them with the theoretical result of Ref. \cite{art05} and the experiment \cite{cha12}.
As one can see from the table, our new result is rather close to the previous theoretical
prediction and disagrees with the experimental result of Ref. \cite{cha12}.

\section{Hyperfine splitting}
\label{sec:2}
High-precision measurements of the hyperfine splitting (HFS) in heavy H-like
ions were performed in Refs. \cite{kla94,cre96,cre98,see98,bei01,ull15}.
The main goal of these experiments was to test QED in a unique combination
of the strongest electric and magnetic fields. For instance, the average magnetic field
experienced by the electron in H-like bismuth amounts to about 30000 T, which is
 1000 times stronger than the field obtained
with the strongest superconducting magnet. 
Despite a very high precision of these measurements, from $10^{-4}$ to $10^{-5}$,
their comparison with theory
could not provide tests of QED because of a large theoretical uncertainty of
the nuclear magnetization distribution correction (so-called Bohr-Weisskopf effect)
\cite{sha97,sen02}.
To overcome this problem, in Ref. \cite{sha01} it was proposed to study a specific difference
of the HFS values for Li- and H-like ions of the same heavy isotope,
\begin{equation}
\Delta'E=\Delta E^{(2s)}-\xi \Delta E^{(1s)}\,,
\end{equation}
where the parameter $\xi$ must be chosen to cancel the Bohr-Weisskopf effect.
It was shown \cite{sha01}  that both the parameter $\xi$ and the specific
difference $\Delta'E$ are very stable with respect to possible variations
of microscopic nuclear models and, therefore, can be calculated to a very
high accuracy. In case of $^{209}$Bi, one finds $\xi = 0.16886$.
The most elaborated theoretical calculations for  $^{209}$Bi employing the nuclear
magnetic moment $\mu/\mu_N=4.1106(2)$ \cite{rag89}
lead  to the specific difference  $\Delta'E=-61.320(6)$ meV \cite{vol12}.
This theoretical value includes the QED contribution of 0.229(2) meV.
It means that tests of QED effects on the HFS are possible, provided
the HFS measurements for H- and Li-like bismuth are carried out to the
required accuracy.

The recent experiments for  $^{209}$Bi \cite{ull17}
resulted in the value $\Delta'E=-61.012 (5)(21)$ meV, which was in strong
disagreement with the theoretical prediction. This disagreement was resolved
in Ref. \cite{skr18}, where new  calculations of the magnetic
shielding constants in  $^{209}$Bi(NO$_3$)$_3$ and   $^{209}$BiF$_6^-$  clearly showed that
the widely used value
for the nuclear magnetic moment, $\mu/\mu_N=4.1106(2)$ \cite{rag89}, is incorrect.
Moreover, these calculations combined with 
new NMR measurements of the nuclear magnetic
moment  in $^{209}$Bi(NO$_3$)$_3$ and  $^{209}$BiF$_6^-$ \cite{skr18}
lead to a value $\mu/\mu_N=4.092(2)$.
With this magnetic  moment,  
the theoretical value of the specific  difference
amounts to $\Delta'E=−61.043(5)(30)$ meV, where the first uncertainty is due to
uncalculated corrections and remaining nuclear effects, while
the second one is due to the uncertainty of the new nuclear
magnetic  moment  value.  This theoretical value
agrees well with the experiment.

The most precise to-date value for the nuclear magnetic moment of  $^{209}$Bi can be
obtained via equating the theoretical
and experimental results on the specific difference. This leads to 
 $\mu/\mu_N=4.0900(15)$  \cite{skr18}. However, new  measurements
of the nuclear magnetic moment as well as the HFS in H- and
Li-like  $^{209}$Bi are needed to provide stringent QED tests.

\section{Bound-electron $g$ factor}
\label{sec:3}
High-precision measurements of the $g$ factor of highly charged ions
were first performed for H-like carbon \cite{haf00}.
The uncertainty of the experimental value obtained in that work was mainly defined
by the uncertainty of the electron mass which was accepted that time.
 Combined with the related advance in  theory of the $g$ factor, 
 which included evaluations of the higher-order relativistic 
 recoil and QED corrections \cite{sha02,yer02}, this resulted in
four-times improvement of the accuracy of the electron mass.
Later, the $g$ factor measurements and the related theoretical predictions were
improved in accuracy and extended to other ions   
(see, e.g., Refs. \cite{pac05,stu13,wag13,stu14,sha15,cza18} and references therein).
As a result, in Refs. \cite{stu14,zat17}
the  precision of the atomic mass of
the electron was further improved by a factor of 13.

Recently, the isotope shift of the $g$ factor of Li-like calcium,
$^{A}$Ca$^{17+}$ with $A=40$ and $A=48$, was measured \cite{koe16}.
From the theoretical side, the value of the isotope shift in calcium
is mainly defined by the nuclear recoil effect (mass shift).
The theoretical result presented in Ref.  \cite{koe16}
included the QED  calculation of the one-electron recoil effect
and the evaluation of the two-electron
recoil effect using an extrapolation of the
results obtained in Refs. \cite{yan01,yan02} in
the framework of a two-component Breit-approximation  approach \cite{heg75}. This theoretical
value was in agreement with the experimental
one but at the edge of the experimental uncertainty. In Ref. \cite{sha17}
the two-electron recoil contribution was recalculated within the Breit
approximation using a four-component approach \cite{sha01b}.
It was found that the obtained result strongly disagrees with
 the corresponding result based on the two-component
 approach  \cite{yan01,yan02}. A detailed analysis \cite{sha17} showed that the disagreement
was caused by omitting some important terms in the calculation scheme
formulated within the two-component approach in Ref. \cite{heg75}.

In Table \ref{tab:4}, the individual contributions to the isotope
shift  of the $g$ factor of Li-like calcium,
$^{40}{\rm Ca}^{17+} -\; ^{48}{\rm Ca}^{17+}$, are presented.
It can be seen that the theoretical and experimental results are in good
agreement with each other. This gives the first test of the relativistic
theory of the recoil effect in highly charged ions in presence of  magnetic field.
As was shown in Ref. \cite{mal17}, the study
of  a specific difference of the $g$ factors of  H- and Li-like lead
 can provide a test of the QED recoil effect on a few-percent level. This
would give the first test of QED at
strong-coupling regime beyond the Furry picture.

\section{Supercritical fields}
\label{sec:4}

\begin{figure}
   \includegraphics[width=1.0\textwidth]{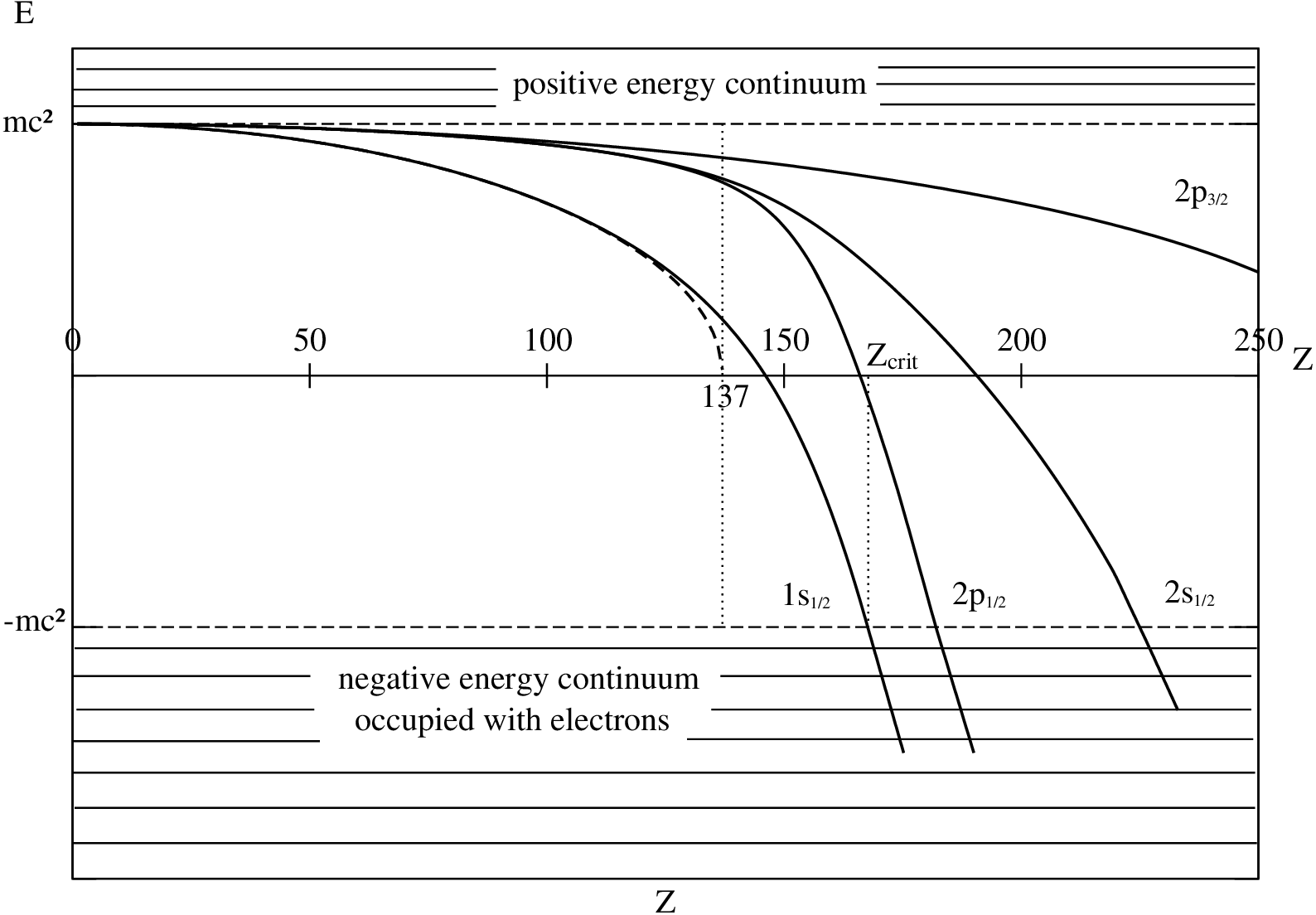}
  \caption{The energy levels of an electron  in a Coulomb field as functions of the nuclear charge number $Z$.
    The dashed line indicates the $1s$ energy for point nuclei. The solid lines respresent
    the energies for extended nuclei.  }
\label{fig:1}       
\end{figure}

In Fig.~\ref{fig:1} we display the energy levels of an electron in a Coulomb field
 as functions of the nuclear charge number $Z$.
For point nuclei the $1s$ level exists only up to  $Z\approx 137$.
However, for extended nuclei
the ground state level goes continuously down and at $Z\approx 173$ 
``dives''\, into the negative-energy 
Dirac continuum (see, e.g.,  book \cite{gre85} and references therein). 
If this level was originally empty, its diving into
the negative-energy Dirac continuum should result in so-called spontaneous emission of two  positrons.
Since there are no nuclei with so high $Z$, the only way  to observe
this fundamental effect is to study low-energy heavy-ion collisions. 

If we consider the behavior of the low-lying energy levels of
a quasimolecule formed by colliding uranium ions
at the energy near the Coulomb  barrier  ($\sim$ 5.9 MeV/u) as functions of time, we find that the supercritical
field exists for about $10^{-21}$ seconds only  \cite{gre85}.
Unfortunately, this time is by two orders of magnitude smaller than the time required
for the spontaneous positron emission, and the positrons are mainly
created due to the dynamical (induced) mechanism.
This was one of the main reasons why the experiments performed at GSI more
than 30 years ago could not prove or disprove the spontaneous positron creation  \cite{gre85}.
One may expect, however, that  investigations of quantum dynamics of
various processes which take place in low-energy collisions of bare nuclei or few-electron
ions  at both subcritical and supercritical regimes can prove
or disprove  the ``diving''\,  scenario which should lead to the spontaneous pair creation.
From the theoretical side, new methods should be developed to perform these investigations in all details.
The calculations of the pair-creation probabilities performed many years ago
by Frankfurt$\,$'s group  \cite{gre85}
were limited by  the monopole approximation.  In this approximation
the expansion of the two-center nuclear potential is restricted to the zero-order spherical harmonic term.
The first calculations  of the pair-creation probabilities
beyond the monopole approximation have been performed in Refs. \cite{mal17a,pop18}.
These calculations at  the energy near the Coulomb  barrier 
showed that the difference between the exact and monopole-approximation results
varies from  about 6\% at the zero impact parameter, $b=0$, to about 30\% at $b=40$ fm.

\section{Conclusion and outlook}
\label{sec:5}

In this paper we have reviewed the recent achievements in the study of QED 
with highly charged ions. To date, strong-field QED has been mainly tested in the Lamb-shift
experiments with heavy ions.
However, during the last years a great progress was made from both experimental and theoretical sides
in investigations of the hyperfine splitting and the  $g$ factor with highly charged ions.
High-precision measurements of the $g$ factor of heavy few-electron ions
are anticipated in the nearest future
at the Max-Planck-Institut f\"ur Kernphysik in Heidelberg
and at the HITRAP/FAIR facilities in Darmstadt. Extention of these measurements to ions with
nonzero nuclear spin could also provide the most precise determination of the nuclear magnetic
moments, which are urgently needed for the HFS studies.
Further developments of the theoretical methods to describe quantum dynamics of electrons
in low-energy heavy-ion collisions are also required. It is expected that investigations
of these collisions can prove or disprove  the “diving”\, scenario which leads to the spontaneous
pair creation.

\section*{Acknowledgements}
This work was supported by the Russian Science Foundation (Grant No. 17-12-01097).
%
%
\begin{table}
\caption{The ground-state Lamb shift in $^{238}$U$^{91+}$, in eV.}
\label{tab:1}       
\begin{tabular}{ll}
\hline\noalign{\smallskip}
Contribution & Value  \\
\noalign{\smallskip}\hline\noalign{\smallskip}
Furry picture QED & 265.19(33)  \\
Finite nuclear size  &  198.54(19)  \\
Nuclear recoil  &  $\;\;\;\,$0.46  \\
Nuclear polarization  & $\;\;\;\,$0.20(10)  \\
Total theory  & 463.99(39)  \\
Experiment \cite{gum05}  &   460.2(4.6) \\
\noalign{\smallskip}\hline
\end{tabular}
\end{table}

\begin{table}
\caption{The $1s2p\,^1P_{1} - 1s^2\,^1S_0$ transition energy in He-like titanium, in eV.}
\label{tab:2}       
\begin{tabular}{ll}
  \hline\noalign{\smallskip}
Contribution & Value  \\
\noalign{\smallskip}\hline\noalign{\smallskip}
Breit approximation  & 4751.706 \\ 
QED  & $\;\;\,-$2.059 \\ 
Total theory (this work)  & 4749.647(2)  \\
Total theory \cite{art05}  & 4749.644(1) \\
Experiment  \cite{cha12}  & 4749.85(7) \\ 
\noalign{\smallskip}\hline
\end{tabular}
\end{table}

\begin{table}
  \caption{
 Isotope shift of the $g$ factor of Li-like calcium,
$^{40}{\rm Ca}^{17+} -\; ^{48}{\rm Ca}^{17+}$.}
\label{tab:4}       
\begin{tabular}{ll}
\hline\noalign{\smallskip}
Contribution & Value  \\
\noalign{\smallskip}\hline\noalign{\smallskip}
   Nuclear recoil: one-electron non-QED                   & $\;\;\,$0.000 000 012 246  \\
           Nuclear recoil: interelectronic int.       &   $-$0.000 000 001 302 \\
Nuclear recoil: QED  &   $\;\;\,$0.000 000 000 114(12) \\
 Finite nuclear size &  $\;\;\,$0.000 000 000 004(9) \\
Total theory \cite{sha17} &  $\;\;\,$0.000 000 011 056(16) \\
Experiment \cite{koe16}  & $\;\;\,$0.000 000 011 7(14)\\
\noalign{\smallskip}\hline
\end{tabular}
\end{table}



\begin{thebibliography}{}
%
  %

\bibitem{gla11}
  Glazov, D.A. {et al.:}
Tests of fundamental theories with heavy ions at low-energy regime.
Hyperfine Interactions {\bf 199}, 71-83 (2011) 
  
\bibitem{koz08}
  Kozhedub, Y.S. {et al.:}
  Nuclear deformation effect on the binding energies in heavy ions.
  Phys. Rev. A {\bf 77}, 032501 (2008)
  
\bibitem{yer15}
 Yerokhin, V.A., Shabaev, V.M.:
Lamb shift of $n = 1$ and $n = 2$ states of hydrogen-like atoms, $1\le  Z \le 110$.
Journal of Physical and Chemical Reference Data {\bf 44}, 033103 (2015) 

\bibitem{yer03}
 Yerokhin, V.A., Indelicato, P., Shabaev, V.M.:
 Evaluation of the two-loop self-energy correction to the ground state energy of H-like ions to all orders in $Z\alpha$.
  Eur. Phys. J.  D {\bf 25}, 203-238 (2003) 

  
\bibitem{sha98}
  Shabaev, V.M.  {et al.:}
Recoil correction to the ground-state energy of hydrogenlike atoms.
Phys. Rev. A {\bf 57}, 4235-4239 (1998)

\bibitem{plu95}
Plunien, G., Soff, G.:
Nuclear-polarization contribution to the Lamb shift in actinide nuclei.
Phys. Rev. A {\bf 51}, 1119-1131 (1995); Erratum: Phys. Rev. A {\bf 53}, 4614 (1996)

\bibitem{nef96}
  Nefiodov, A.V.  {et al.:}
Nuclear polarization effects in spectra of multicharged ions.
Phys. Lett. A {\bf 222},  227-232  (1996)

\bibitem{gum05}
  Gumberidze, A.  {et al.:}
Quantum electrodynamics in strong electric fields: the ground-state Lamb shift in hydrogenlike uranium.
Phys. Rev. Lett. {\bf 94}, 223001 (2005)


\bibitem{sch91}
  Schweppe, J.  {et al.:}
Measurement of the Lamb shift in lithiumlike uranium (U$^{89+}$).
Phys. Rev. Lett. {\bf 66}, 1434-1437 (1991)

\bibitem{bra03}
  Brandau, C.  {et al.:}
  Precise determination of the $2s_{1/2}-2p_{1/2}$
  splitting in very heavy lithiumlike ions utilizing dielectronic recombination.
Phys. Rev. Lett. {\bf 91}, 073202 (2003)

\bibitem{bei05}
  Beiersdorfer, P.  {et al.:}
Measurement of the two-loop Lamb shift in lithiumlike U$^{89+}$.
Phys. Rev. Lett. {\bf 95}, 233003 (2005) 

\bibitem{yer07}
Yerokhin, V.A., Artemyev, A.N., Shabaev, V.M.:
QED treatment of electron correlation in Li-like ions.
Phys. Rev. A {\bf 75}, 062501 (2007) 

\bibitem{sap11}
Sapirstein, J., Cheng, K.T.:
S-matrix calculations of energy levels of the lithium isoelectronic sequence.
Phys. Rev. A {\bf 83}, 012504 (2011) 

\bibitem{cha12}
  Chantler, C.T.  {et al.:}
  Testing three-body quantum electrodynamics with trapped Ti$^{20+}$ ions:
  Evidence for a $Z$-dependent divergence between experiment and calculation.
Phys. Rev. Lett. {\bf 109}, 153001  (2012) 


\bibitem{tra09}
  Trassinelli, M.  {et al.:}
Observation of the $2p_{(3/2)} \rightarrow 2s_{(1/2)}$ intra-shell transition in He-like uranium.
Eur. Phys. Lett. {\bf 87}, 63001 (2009)

\bibitem{ama12}
  Amaro, P.  {et al.:}
Absolute measurement of the relativistic magnetic dipole transition energy in heliumlike argon.
Phys. Rev. Lett. {\bf 109}, 043005 (2012)

\bibitem{kub14}
  Kubicek, K.  {et al.:}
Transition energy measurements in hydrogenlike and heliumlike ions strongly supporting
bound-state QED calculations.
Phys. Rev. A {\bf 90}, 032508 (2014)

\bibitem{bei15}
Beiersdorfer, P., Brown, G.V.:
Experimental study of the x-ray transitions in the heliumlike isoelectronic sequence: Updated results.
Phys. Rev. A {\bf 91}, 032514 (2015)

\bibitem{epp15}
  Epp, S.W.  {et al.:}
Single-photon excitation of K$\alpha$ in heliumlike Kr$^{34+}$: Results supporting quantum electrodynamics predictions.
  Phys. Rev. A {\bf 92}, 020502(R) (2015)

\bibitem{mac18}
  Machado, J.  {et al.:}
  High-precision measurements of $n=2 \rightarrow n=1$
  transition energies and level widths in He- and Be-like argon ions.
 Phys. Rev. A {\bf 97}, 032517 (2018) 
  
\bibitem{art05}
  Artemyev, A.N.  {et al.:}
QED calculation of the $n=1$ and $n=2$ energy levels in He-like ions.
Phys. Rev. A {\bf 71}, 062104 (2005)


\bibitem{kla94}
  Klaft, I.  {et al.:}
Precision laser spectroscopy of the ground state hyperfine splitting of hydrogenlike $^{209}$Bi$^{82+}$.
Phys. Rev. Lett. {\bf 73}, 2425-2427 (1994)

\bibitem{cre96}
  Crespo Lopez-Urrutia, J.R.  {et al.:}
Direct observation of the spontaneous emission of
the hyperfine transition $F=4$ to $F=3$ in ground state hydrogenlike $^{165}$Ho$^{66+}$ in an electron beam ion trap.
{Phys. Rev. Lett.} {\bf 77}, 826-829 (1996)

\bibitem{cre98}
  Crespo Lopez-Urrutia, J.R.  {et al.:}
Nuclear magnetization distribution radii determined
by hyperfine transitions in the $1s$ level of H-like ions $^{185}$Re$^{74+}$ and $^{187}$Re$^{74+}$.
{Phys. Rev. A} {\bf 57}, 879-887 (1998)

\bibitem{see98}
  Seelig, P.  {et al.:}
 Ground state hyperfine splitting of hydrogenlike $^{207}$Pb$^{81+}$ by laser excitation
 of a bunched ion beam in the GSI experimental storage ring.
 {Phys. Rev. Lett.} {\bf 81}, 4824-4827 (1998) 

\bibitem{bei01}
  Beiersdorfer, P.  {et al.:}
Hyperfine structure of hydrogenlike thallium isotopes.
Phys. Rev. A {\bf 64}, 032506 (2001)

\bibitem{ull15}
  Ullmann, J.  {et al.:}
  An improved value for the hyperfine splitting of hydrogen-like $^{209}$Bi$^{82+}$.
  J. Phys. B {\bf 48}, 144022  (2015) 

\bibitem{sha97}
  Shabaev, V.M.  {et al.:}
Ground-state hyperfine splitting of high-$Z$ hydrogenlike ions.
Phys. Rev. A {\bf 56}, 252-255 (1997) 

\bibitem{sen02}
Sen'kov, R.A., Dmitriev, V.F.:
Nuclear magnetization distribution and hyperfine splitting in Bi$^{82+}$ ion.
Nucl. Phys. A {\bf 706}, 351-364  (2002) 

\bibitem{sha01}
  Shabaev, V.M.  {et al.:}
  Towards a test of QED in investigations of the hyperfine splitting in heavy ions.
Phys. Rev. Lett. {\bf 86}, 3959-3962   (2001) 

  \bibitem{rag89}
    Raghavan, P.:
Table of nuclear moments.    
  {At. Data Nucl. Data Tables} {\bf 42}, 189-291 (1989)

\bibitem{vol12}
  Volotka, A.V.  {et al.:}
  Test of many-electron QED effects in the hyperfine splitting of heavy high-$Z$ Ions.
  Phys. Rev. Lett. {\bf 108}, 073001  (2012) 
  
\bibitem{ull17}
  Ullmann, J.  {et al.:}
  High precision hyperfine measurements in bismuth challenge bound-state strong-field QED.
Nature Communications {\bf 8}, 15484  (2017)

  
\bibitem{skr18}
  Skripnikov, L.V.  {et al.:}
  New nuclear magnetic moment of $^{209}$Bi: Resolving the bismuth hyperfine puzzle.
  Phys. Rev. Lett. {\bf 120}, 093001  (2018) 

\bibitem{haf00}  
  H\"affner, H.  {et al.:}
High-accuracy measurement of the magnetic moment anomaly of the electron bound in hydrogenlike carbon.
Phys. Rev. Lett. {\bf 85}, 5308-5311 (2000) 

\bibitem{sha02}
Shabaev, V.M., Yerokhin, V.A.:
Recoil correction to the bound-electron $g$ factor in H-like atoms to all orders in $\alpha Z$.
Phys. Rev. Lett. {\bf 88}, 091801 (2002)

\bibitem{yer02}
Yerokhin, V.A.,  Indelicato, P.,  Shabaev, V.M.:
Self-energy correction to the bound-electron $g$ factor in H-like ions.
Phys. Rev. Lett. {\bf 89}, 143001 (2002)


\bibitem{pac05}
  Pachucki, K.  {et al.:}
Complete two-loop correction to the bound-electron $g$ factor.
Phys. Rev. A {\bf 72}, 022108 (2005) 

\bibitem{stu13}
Sturm, S., Werth, G., Blaum, K.:
Electron $g$-factor determinations in Penning traps.
Annalen Der Physik {\bf 525}, 620-635 (2013)

\bibitem{wag13}
  Wagner, A.  {et al.:}
  $g$ factor of lithiumlike silicon $^{28}$Si$^{11+}$.
Phys. Rev. Lett. {\bf 110}, 033003 (2013)

\bibitem{stu14}
  Sturm, S.  {et al.:}
High-precision measurement of the atomic mass of the electron.
Nature  {\bf 506}, 467-470 (2014)

\bibitem{sha15}
  Shabaev, V.M.  {et al.:}
Theory of bound-electron $g$ factor in highly charged ions.
Journal of Physical and Chemical Reference Data {\bf 44}, 031205 (2015) 


\bibitem{cza18}
  Czarnecki, A.  {et al.:}
Two-loop binding corrections to the electron gyromagnetic factor.
Phys. Rev. Lett. {\bf 120}, 043203 (2018)


\bibitem{zat17}
  Zatorski, J.  {et al.:}
Extraction of the electron mass from $g$-factor measurements on light hydrogenlike ions.
Phys. Rev. A {\bf 96}, 012502 (2017)

\bibitem{koe16}
  K\"ohler, F.  {et al.:}
Isotope dependence of the Zeeman effect in lithium-like calcium.
Nature Communications {\bf 7}, 10246 (2016) 


\bibitem{yan01}
 Yan, Z.-C.:
Calculations of magnetic moments for three-electron atomic systems.
Phys. Rev. Lett. {\bf 86}, 5683-5686 (2001)

\bibitem{yan02}
 Yan, Z.-C.:
Calculations of magnetic moments for lithium-like ions.
J. Phys. B {\bf 35}, 1885-1892 (2002)

\bibitem{heg75}
Hegstrom, R.A.:
Magnetic moment  of atomic lithium.
Phys. Rev. A {\bf 11}, 421-426 (1975).

\bibitem{sha17}
  Shabaev, V.M.  {et al.:}
Recoil effect on the $g$ factor of Li-like ions.
Phys. Rev. Lett. {\bf 119}, 263001 (2017)


\bibitem{sha01b}
  Shabaev, V.M.:
QED theory of the nuclear recoil effect on the atomic $g$ factor.
Phys. Rev. A {\bf 64}, 052104 (2001) 

\bibitem{mal17}
  Malyshev, A.V.  {et al.:}
Nuclear recoil effect on the $g$-factor of heavy ions:
Prospects for tests of quantum electrodynamics in a new region.
JETP Letters  {\bf 106}, 765-770  (2017) 

\bibitem{gre85}
  Greiner, W., Muller, B., Rafelski, J.,
  Quantum electrodynamics of strong fields, 
  Springer-Verlag, Berlin  (1985)

  \bibitem{mal17a}
    Maltsev, I.A.  {et al.:}
Pair production in low-energy collisions of uranium nuclei beyond the monopole approximation.
Nucl. Instr. Methods Phys. Res. B {\bf 408}, 97-99 (2017) 
  
\bibitem{pop18}
  Popov, R.V.  {et al.:}
   One-center calculations of the electron-positron pair creation in low-energy collisions of heavy bare nuclei.
    Eur. Phys. J. D {\bf 72}, 115 (2018) 

  



\end{thebibliography}


\end{document}